\documentclass[superscriptaddress,twocolumn,showkeys,amsmath,amssymb,prl,showpacs]{revtex4}

\newcommand{\eref}[1]{eq.~(\ref{#1})}

\newcommand{\fref}[1]{fig.~\ref{#1}}

\newcommand{\Eref}[1]{Eq.~(\ref{#1})}

\newcommand{\letter}[1]{\title{#1}}
\newcommand{\ead}[1]{\email{#1}}
\newcommand{\submitto}[1]{}
\newcommand{\rmd}{\mathrm{d}}
\newcommand{\rme}{\mathrm{e}}

\newcommand{\Tr}{\mathop{\mathrm{Tr}}\nolimits}
\newcommand{\tr}{\mathop{\mathrm{tr}}\nolimits}
\newcommand{\Or}{\mathord{\mathrm{O}}} 

\renewcommand{\address}{\affiliation}

\usepackage{graphicx}
\graphicspath{{../tesis/eps/}{eps/}}
\def\>{\rangle} \def\<{\langle}
\newcommand{\qm}{\mathrm{qm}}
\newcommand{\e}{\rme}
\newcommand{\into}{\int_0^t \rmd \tau \int_0^t \rmd \tau'}
\newcommand{\ie}{\textit{i.e.} }

\newcommand{\intoh}{\int_0^t \rmd \tau \int_0^\tau \rmd \tau'}
\renewcommand{\section}[1]{}
\renewcommand{\subsection}[1]{}
\newcommand{\He}{H}
\renewcommand{\H}{\mathcal H}
\begin{document}
\letter{Decoherence of an $n$-qubit quantum memory}

\newcommand{\unam}{Universidad Nacional Aut\'onoma de M\'exico, M\'exico}
\newcommand{\ifunam}{\address{Instituto de F\'{\i}sica, \unam}}
\newcommand{\fisguadalajara}{\address{Departamento de F\'isica, Universidad de Guadalajara, M\'exico}}
\newcommand{\cic}{\address{Centro Internacional de Ciencias, Cuernavaca, M\'exico}}
\newcommand{\icf}{\address{Instituto de Ciencias F\'{\i}sicas, \unam}}

\author{Thomas Gorin}  \fisguadalajara \cic
\author{Carlos Pineda} \ead{carlospgmat03@gmail.com} \cic \icf \ifunam
\author{Thomas H. Seligman} \cic \icf

\date{\today}

\begin{abstract}
We analyze decoherence of a quantum register in the absence of non-local
operations  \ie of $n$ non-interacting qubits coupled to an
environment.  The problem is solved in terms of a sum rule which implies linear
scaling in the number of qubits.  Each term involves a single qubit and its
entanglement with the remaining ones.  Two conditions are essential: first
decoherence must be small and second the coupling of different qubits must be
uncorrelated in the interaction picture.  We apply the result to a random
matrix model, and illustrate its reach considering a GHZ state coupled to a
spin bath. 
\end{abstract}
\pacs{03.65.-w,03.65.Yz,05.40.-a}
\keywords{quantum memory, quantum register, decoherence, GHZ, entanglement, purity}

\maketitle

Physical devices capable of storing faithfully a quantum state, \ie quantum
memories, are crucial for any quantum information task.  While different types
of systems have been considered, most of the effort is concentrated in
manipulating qubits as they are essential for most quantum information
tasks \cite{NC00a}.  For a general quantum memory (QM) it is necessary to
record, store and retrieve an arbitrary state. For quantum communication and
quantum computation initialization in the base state, single qubit
manipulations, a two qubit gate (e.g. control-not), and single qubit
measurements are sufficient. For arbitrary quantum memories recording and
retrieving the state is still an experimental challenge but initializing a
qubit register and individual measurement is mostly mastered.
Faithful realization of two qubit gates together with better
isolation from the environment are the remaining obstacles for achieving fully
operational quantum technology.  However the huge effort done in the  community
has born some fruit \cite{comercialQI, qmemoryexpAPS}.

A number of papers discuss how two qubit quantum memories are
affected by coupling to some environment. In our previous work we
study the case of two qubits interacting either with a spin bath
\cite{pineda:012305} or a random matrix environment \cite{pinedaRMTshort,pinedalong}. In
both cases the loss of coherence and the loss of internal entanglement of the
pair was studied. Many other papers focus on the loss of internal entanglement
only, due to the interaction with a thermal bath (see \cite{andrereview}
and references therein).  A study of decoherence of a two-qubit system as
measured by entanglement fidelity is presented in \cite{QMstoragedecoherenceetal}.
There the effect of fluctuations of the parameters of the Hamiltonian is
analyzed.  In \cite{1996RSPSA.452..567:M,quantumregisterdecoherence} an 
$n$ qubit QM coupled to a bath of harmonic oscillators is studied.

In the present letter we  give analytic expressions for the decoherence of a QM
during the storage time.  Specifically we discuss a QM composed of
a set of individual qubits interacting with an environment. The
expressions given are based on previous knowledge of
the decoherence of a single qubit entangled with some spectator, with which it
does not interact. Their validity is limited to small decoherence, \ie
large purity of the QM. Note that the latter is no significant restriction due
to the high fidelity requirements for the application of
quantum error correction codes.  We assume that the
entire system is subject to unitary time-evolution, and that decoherence comes
about by entanglement between the central system and the environment. Because
of its simple analytical structure decoherence shall be measured in terms of
purity \cite{Zur91} of the central system.   Spurious interactions inside the
central system are neglected.  We shall rely heavily on our recent studies of
decoherence of two qubits \cite{pinedaRMTshort, pinedalong}.

A further and critical assumption is the independence of the coupling in the
interaction picture. This is justified if the couplings are independent in the
Schr\"odinger picture or if we have rapidly decaying correlations due to mixing
properties of the environment in the spirit of \cite{purityfidelity}.
Physically the first would be more likely if we talk about qubits realized in
different systems or degrees of freedom, while the second seems plausible for
many typical environments.

We generalize the concept of a spectator configuration \cite{pineda:012305,
pinedaRMTshort, pinedalong}: The central system consists of two non-interacting
parts, one interacting with the environment and the other not. This configuration 
is non-trivial if the two parts of the central system are entangled. In our case
the first part will be a single qubit and the second part (the spectator)
will consist of all other qubits of the QM. Our central result is then 
a decomposition of the decoherence of the full QM, as measured by purity,
in a sum of terms describing the decoherence of each qubit
in a spectator configuration. Apart from the above made assumptions,
this result does not depend on any particular property of the environment
or the coupling.  Thus it can be applied to a variety of models.

We test successfully the results in a random matrix model both for coupling and bath
\cite{pinedaRMTshort, pinedalong}.  The general relation is obtained in linear
response approximation and leads to explicit analytic results if the spectral
correlations of the bath are known.  Finally, we perform numerical simulations for
two and four qubits, interacting with a kicked Ising spin chain \cite{pineda:012305, 
prosenKI, pp2007} as an environment, which can be chosen mixing or integrable. 

The central system (our QM) is composed of $n$ qubits. Thus its Hilbert space 
is $\H_\qm=\bigotimes_{i=1}^{n}\H_i$, where $\H_i$ are the Hilbert spaces of 
the qubits ($\dim \H_i=2$). The Hilbert space of the environment is denoted by 
$\H_\e$. The Hamiltonian reads as 
\begin{equation}
H= H_\qm + H_\e + \lambda V\; .
\label{eq:fullham}
\end{equation}
Here, $H_\qm=\sum_{i=1}^{n} H_i$, where $H_i$ acts on $\H_i$, whereas 
$H_\e$ describes the dynamics of the environment and 
\begin{equation}
\lambda V=\sum_{i=1}^{n} \lambda_i V^{(i)}
\label{eq:couplings}
\end{equation}
the coupling of the qubits to the environment. The strength of the
coupling of qubit $i$ is controlled by the parameter $\lambda_i$, while 
$\lambda=\max \{|\lambda_i|\}$.

We consider two different settings. In the first one $V^{(i)}$ acts on the
space $\H_i\otimes \H_\e$, \ie all qubits 
interact with a common environment. We shall name this the joint
environment configuration.  In the second one each qubit interacts with a
separate environment. Thus the environment is split into $n$ parts,
$\H_\e=\bigotimes_{i=1}^{n}\H_{\e,i}$ and $V^{(i)}$, in
\eref{eq:couplings}, acts only on $\H_i\otimes \H_{\e,i}$. This we call the
separate environment configuration. The first case would be typical for a quantum
computer, where all qubits are close two each other, while the second
would apply to a non-local quantum network.

We choose the initial state to be the product of a pure state of the 
central system (the QM) and a pure state of the environment. It
may be written as
\begin{equation}
      |\psi_0\>=|\psi_\qm \> |\psi_\e \>,\quad |\psi_\qm \>\in \H_\qm,\, |\psi_\e \>\in \H_\e.
      \label{eq:initialstate}
\end{equation}
Thus, the reduced initial state in the quantum memory $\varrho_0=\tr_\e |\psi_0 \>\<
\psi_0|=|\psi_\qm \>\< \psi_\qm|$ is pure.  In the separate environment
configuration we furthermore assume that $|\psi_\e \>=\bigotimes_i |\psi_{\e,i }\>$
with $|\psi_{\e,i} \> \in \H_{\e,i}$, which corresponds to the absence of quantum
correlations among the different environments.

For reasons of analytic simplicity we chose the purity $P(t)$ as measure for decoherence.
It is defined as
\begin{equation} \label{eq:purityevolution}
      P(t)=\Tr \rho_\qm ^2(t),\; \rho_\qm(t)=
      \Tr_\e \e^{-\imath H t} |\psi_0\> \<\psi_0| \e^{\imath H t}.
\end{equation}
To calculate purity or any other measure of decoherence we can replace the forward
time evolution $\exp(-\imath H t)$ by the echo operator 
$M(t)=\exp(\imath H_0 t)\, \exp(-\imath H t)$ where $H_0$ determines the evolution without 
coupling ($\lambda=0$).
In linear response approximation this operator reads as
\begin{equation}
       M(t)= \openone -\imath \lambda I(t) - \lambda^2 J(t),
       \label{eq:born}
\end{equation}
where $I(t)= \int_0^t \rmd \tau \tilde V_\tau$ and $J(t)= \intoh \tilde V_\tau
\tilde V_{\tau'}$.  $\lambda \tilde V_t = \exp(\imath H_0 t)\, \lambda V  \exp(-\imath H_0 t)$  
is the coupling at time $t$ in the
interaction picture.  As discussed in \cite{pinedalong} it is convenient
to introduce the form
$p [ \rho_1 \otimes \rho_2]= \Tr( \Tr_\e  \rho_1 \Tr_\e \rho_2)$.  
Purity is then given by
\begin{equation}\label{eq:purityA}
P(t)=1-2\lambda^2 \into {\rm Re}\, A(\tau,\tau')+\Or\left(\lambda^3\right)
\end{equation}
with
\begin{multline}\label{eq:AIJdos} 
A(\tau,\tau')= p[\tilde V_\tau \tilde V_{\tau'} \varrho_0\otimes\varrho_0]
  - p[\tilde V_{\tau'} \varrho_0 \tilde V_\tau \otimes\varrho_0]  \\
  + p[\tilde V_\tau \varrho_0\otimes \tilde V_{\tau'} \varrho_0]
  - p[\tilde V_{\tau'} \varrho_0\otimes \varrho_0 \tilde V_\tau] \; .
\end{multline}
Note that the linear terms vanish after tracing out the environment. 
Considering the form of the coupling in \Eref{eq:couplings} we can decompose 
$\lambda^2 A(\tau,\tau')=\sum_{i,j} \lambda_i \lambda_j A^{(i,j)}(\tau,\tau')$
with 
\begin{multline}\label{eq:AIJdecom} 
A^{(i,j)}(\tau,\tau')=
    p[\tilde V^{(i)}_\tau \tilde V^{(j)}_{\tau'} \varrho_0\otimes\varrho_0]
  - p[\tilde V^{(i)}_{\tau'} \varrho_0 \tilde V^{(j)}_\tau \otimes\varrho_i]\\
  + p[\tilde V^{(i)}_\tau \varrho_0\otimes \tilde V^{(j)}_{\tau'} \varrho_0]
  - p[\tilde V^{(i)}_{\tau'} \varrho_0\otimes \varrho_0 \tilde V^{(j)}_\tau]
\; . 
\end{multline}
$\tilde V^{(i)}_t= \exp(\imath H_0 t)\, V^{(i)}  \exp(-\imath H_0 t)$
in analogy with $\tilde V_t$. 
Equation~(\ref{eq:purityA}) is then given as a double sum in the indices of 
the qubits.  In the diagonal approximation ($i=j$), $P(t)$ can be expressed in 
terms of purities $P^{(i)}_{\rm (sp)}(t)$ associated with individual qubits
\begin{eqnarray}\label{eq:spectatorpurity}
       P(t)&=& 1- \sum_{i=1}^n\left( 1- P^{(i)}_{\rm (sp)}(t) \right),\\
	P^{(i)}_{\rm (sp)}(t)&=& 1- 2 \lambda_i^2\into  A^{(i,i)} (\tau,\tau') 
      + \Or\left(\lambda_i^3\right).\nonumber
\end{eqnarray}
Each contribution corresponds to the purity decay of 
the central system in a spectator configuration with qubit 
$i$ interacting with the environment.
\Eref{eq:spectatorpurity} is the central result of this paper. The 
diagonal approximation is justified in two 
situations: First if the couplings $V^{(i)}$ of the individual 
qubits are independent from the outset, as would be typical for the separate 
environment configuration or for the random matrix model of decoherence 
\cite{pinedalong}. Second, if the couplings in the interaction picture become 
independent due to mixing properties of the environment, as would be typical 
for a ``quantum chaotic'' environment.

To illustrate the case of independent couplings mentioned above, random matrix
theory provides a handy example. Such models were discussed in
\cite{1464-4266-4-4-325,pinedaRMTshort,otroRMT} and describe the couplings $V^{(i)}$ by 
independent random matrices, chosen from the classical 
ensembles~\cite{classicalensemblesAPS}.

In \cite{pinedalong,pinedaRMTshort} the purity decay was computed in linear
response approximation for two qubits, one of them being the spectator. For the
sake of simplicity we choose the joint environment configuration, no internal
dynamics for the qubits, and let $H_\e$ and $V^{(1)}$ be typical members of the
Gaussian unitary ensemble. Purity is then given by
\begin{eqnarray}
P^{(1)}_{\rm (sp)}(t) &=& 1-\lambda_1^2 (2-p_1) f(t); \label{eq:gue}\\
f(t) &=& t\, \max \{t,\tau_H \} 
   + \frac{2}{3\tau_H}\, \min \{ t,\tau_H\}^3.
\end{eqnarray}
Here, $\tau_\He$ is the Heisenberg time of the environment, and $p_1$ is the 
{\it initial} purity of the first qubit alone, which measures its entanglement 
with the rest of the QM. As we required independence of the couplings, we can 
now insert \eref{eq:gue} in \eref{eq:spectatorpurity} to obtain the simple
expression
\begin{equation}\label{eq:sepgen}
       P(t)= 1-f(t) \sum_{i=1}^{n}\lambda_i^2 (2-p_i),
\end{equation}
where $p_i$ is the initial purity of qubit $i$. In the presence of internal
dynamics the spectator result is also known~\cite{pinedalong} and can be 
inserted. As an example, we apply the above equation to an initial GHZ 
state ($(|0\cdots0\>+ |1\cdots1\>)/\sqrt{2}$). Then all $p_i=1/2$ and we obtain 
$P(t)=1-(3/2)f(t)\sum_i \lambda_i^2$. For a W state
$(|10\ldots0\>+|01\ldots0\>+\cdots+|00\ldots1\>)/\sqrt{n}$,
purity for each qubit is $p_i=(n^2-2n+2)/n^2 \approx 1$, in the
large $n$ limit, and purity decays as $P(t)=1-f(t)\sum_i \lambda_i^2$.
For $n=2$ we retrieve the results published in \cite{pinedalong}.

\begin{figure}
\includegraphics{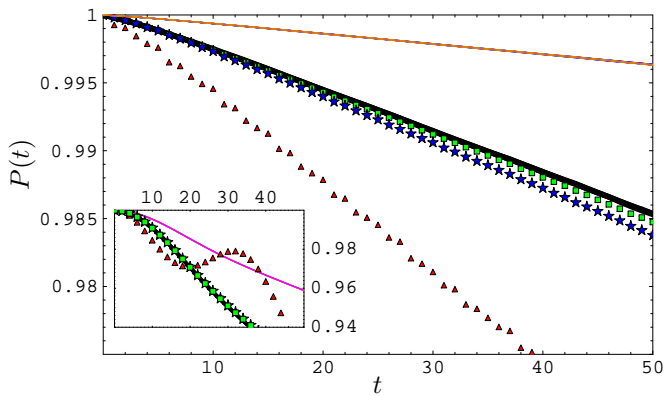}
\caption{(Color online) Purity decay for a GHZ state is shown for the environment
	described by \eref{eq:hamkie}. The couplings are:
	all qubits to one spin (red triangles), nearby spins (blue stars),
	and maximally separated spins (green squares). The theoretical result
	(thick line) is calculated in terms of the $P^{(i)}_{\rm (sp)}(t)$
	(thin lines). The main figure corresponds to the case of a mixing environment
	and the inset for an integrable one.  }
\label{fig:kichaos}
\end{figure}

The main assumption in \Eref{eq:spectatorpurity} is the fast decay of correlations
for couplings of different qubits to the bath. For the random matrix model
discussed above this is trivially fulfilled. Yet, integrable environments are
commonly used~\cite{caldeiraleggett} and one may wonder whether
\Eref{eq:spectatorpurity} can hold in such a context. We shall therefore study a
dynamical model where a few qubits are coupled to an environment represented by
a kicked Ising spin chain using identical coupling operators for all qubits.
In this model, the variation of the angle of the external kicking field allows
the transition from a ``quantum chaotic'' to an integrable Hamiltonian for the
environment~\cite{prosenKI}. The Hamiltonian of the chain is given by
\begin{equation}
H^{(\e)}= \sum_{i=0}^{L-1}  \sigma_x^{(\e,i)} \sigma_x^{(\e,i+1)}
  +\boldsymbol{\delta}(t) \sum_{i=0}^{L-1} b \cdot \sigma^{(\e,i)} 
\label{eq:hamkie}
\end{equation}
where $\boldsymbol{\delta}(t)$
is a train of Dirac delta functions at integer times, $L$ 
is the number of spins in the environment, 
$\sigma^{(\e,i)}=(\sigma_x^{(\e,i)},\sigma_y^{(\e,i)},\sigma_z^{(\e,i)})$ are
the Pauli matrices of spin $i$, and $b$ the dimensionless magnetic field
with which the chain is kicked ($\hbar=1$). We close the ring requiring 
$\sigma^{(\e,L)} =\sigma^{(\e,0)}$. The Hamiltonian of the QM is 
$H^{(\qm)}=\boldsymbol{\delta}(t) \sum_i  b \cdot \sigma^{(\qm,i)}$,
where $\sigma^{(\qm,i)}$ is defined similarly as for the 
environment. The coupling is given by
$\lambda V=\lambda \sum_i \sigma_x^{(\qm,i)} \sigma_x^{(\e,j_i)}$, where the
$j_i$'s define the positions where the qubits are coupled to the spin chain,
e.g. if all $j_i$ are equal, all qubits of our QM are coupled to the same 
position. 

To implement a chaotic environment, we use $b=(0.9,0.9,0)$.
For these values the quantum-chaos properties have
been tested extensively~\cite{pp2007}. We use a ring consisting of $L=12$ spins
for the environment and $4$ additional spins for the qubits of the QM. The 
coupling strength has been fixed to $\lambda=0.005$. In 
Fig.~\ref{fig:kichaos}, we study purity decay when all four qubits are coupled
to the same spin, when they are coupled to neighboring
spins, and when they are coupled to maximally separated
spins, $\vec j=(0,3,6,9)$. The initial state is the product of a GHZ state
in the QM and a random pure state in the environment. We compare the results
with the prediction of \Eref{eq:spectatorpurity} obtained from simulations of the
corresponding spectator configurations (thin solid line). All spectator
configurations yield almost the same result for $P(t)$, so that we can see only
one line. The figure demonstrates the validity of 
\eref{eq:spectatorpurity} for well separated and hence independent couplings. For
smaller separations we obtain faster decay.

Next we perform similar calculations for integrable environments. A 
typical case is shown in the inset of \fref{fig:kichaos} for a two  qubit QM
and $b=(0,1.53,0)$. Both qubits are coupled to the same, neighboring or opposite
spins in the ring. Purity decay is faster than in the mixing case as can be expected
from general considerations in \cite{reflosch}. 
Nevertheless the sum rule is again well fulfilled except if we couple both qubits
to the same spin. 
This leads us to check the
behavior of the correlation function
\begin{equation}\label{eq:domterm}
p[\tilde V^{(i)}_\tau \tilde V^{(j)}_{\tau'}
\varrho_0\otimes\varrho_0] = \<\psi_0|\tilde V^{(i)}_\tau \tilde
V^{(j)}_{\tau'} |\psi_0\>.
\end{equation}
This is the simplest and usually largest term in
$A^{(i,j)}$, \eref{eq:AIJdecom}. Figure \ref{fig:kicorrel} shows this quantity for mixing
and integrable environments in the first and second row respectively.  The first
column shows the autocorrelation function ($i=j$). The second and third column
give the cross correlation function ($i \ne j$) when the qubits are coupled to
the same or opposite spins respectively.  We see that in the latter case correlations 
in both the integrable and the chaotic case are small, 
thus showing that also in integrable situations our condition
can be met.
Non-vanishing correlations, as in \fref{fig:kicorrel}(b) and \fref{fig:kicorrel}(e)
lead to deviations from the sum rule. The pronounced structure 
for the integrable environment, \fref{fig:kicorrel}(e), may be associated 
with the oscillations of purity decay (inset of \fref{fig:kichaos}). 
\begin{figure}
\includegraphics[width=\columnwidth]{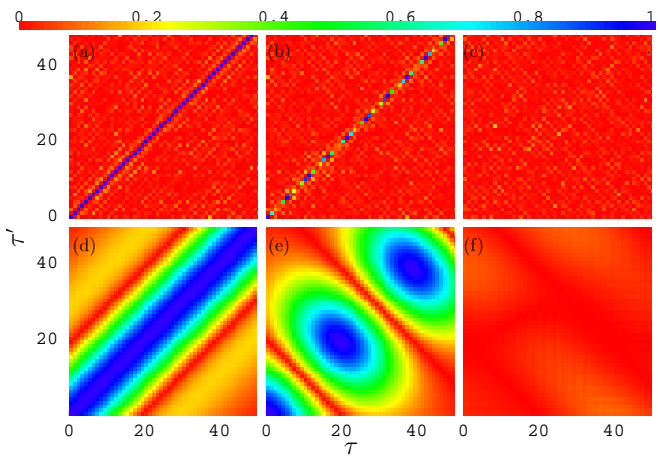}
\caption{(Color online) The color-coded absolute value of the real part of 
	$\<\psi_0|\tilde V^{(i)}_\tau \tilde V^{(j)}_{\tau'} |\psi_0\>$ given
	in \eref{eq:domterm} as a function of $\tau$ and $\tau'$ for the
	integrable [(a) (b) (c)] and the chaotic [(d) (e) (f)] spin chain
	environment ($L=8$).  (a),(d) show the autocorrelation function
	($i=j$); (b), (e) and (c) (f) the cross correlation function for
	couplings to the same spin or opposite spins in the ring respectively.
	}
\label{fig:kicorrel}
\end{figure}

We have calculated generic decoherence  of an
$n$-qubit quantum memory as represented by purity decay. For large purities it
is globally given  in linear response approximation by a sum rule in terms of the
purities of the individual qubits entangled with the remaining register as
spectator. This sum rule depends crucially on the absence or rapid decay of
correlations between the couplings of different qubits in the interaction
picture. We prove, that this is fulfilled by sufficiently chaotic environments
or couplings, but using a spin chain model as an environment we find, that even
if the latter is integrable correlations can be absent and the sum rule holds.
While exceptions, no doubt, can be constructed we have a very general
tool to reduce the decoherence of a QM of $n$ qubits to the problem of a single
qubit entangled with the rest of the QM. Furthermore the result is equally
valid if we perform local operations on the qubits. The sum rule also applies if the
register is split into arbitrary sets of qubits. In fact one and two qubit
gates are known to be universal for quantum computation \cite{NC00a}, so we lay
the foundation for the computation of decoherence during the execution of a
general algorithm.  This extension only requires the knowledge of the
decoherence of a pair suffering the gate operation while entangled with the
rest of the register. As each gate is different, we will have to take a step by
step approach, which at each step will only involve one and two qubit
decoherence. The linear response approximation will be sufficient due to the
high fidelity requirements of quantum computation.  

\begin{acknowledgments}
  We thank Fran\c{c}ois Leyvraz, Toma\v z Prosen,
  and Stefan Mossmann for many useful discussions.  We
  acknowledge support from the grants UNAM-PAPIIT IN112507 and CONACyT 44020-F.
  C.P. was supported by  VCCIC.
\end{acknowledgments}

\bibliographystyle{apsrev}
\bibliography{paperdef,specialbib,miblibliografia}

\end{document}